\documentclass[twocolumn,english,superscriptaddress,floatfix]{revtex4-2}
\usepackage[T1]{fontenc}
\usepackage[utf8]{inputenc}
\usepackage{amsmath}
\usepackage{graphicx}
\usepackage{babel}
\begin{document}
\title{Surfing, sweeping, and assembly of particles by a moving liquid crystal
phase boundary}
\author{Tom Shneer}
\affiliation{Department of Physics \& Astronomy, Tufts University, 574 Boston Ave, Medford, MA 02155}
\author{Jocelyn Ochoa}
\affiliation{Department of Chemistry \& Biochemistry, University of California, Merced, 5200 N. Lake Merced, CA 95343}
\author{Alauna C. Wheeler}
\affiliation{Department of Physics, University of California, Merced, 5200 N. Lake Merced, CA 953433}
\author{Isabella C. Reyes}
\affiliation{Department of Chemistry \& Biochemistry, Santa Clara University, 500 El Camino Real, Santa Clara, CA 95053}
\author{Chaitanya Joshi}
\affiliation{Department of Physics \& Astronomy, Tufts University, 574 Boston Ave, Medford, MA 02155}
\author{Benjamin J. Stokes}
\affiliation{Department of Chemistry \& Biochemistry, Santa Clara University, 500 El Camino Real, Santa Clara, CA 95053}
\author{Linda S. Hirst}
\affiliation{Department of Physics, University of California, Merced, 5200 N. Lake Merced, CA 953433}
\author{Timothy J. Atherton}
\email{timothy.atherton@tufts.edu}
\affiliation{Department of Physics \& Astronomy, Tufts University, 574 Boston Ave, Medford, MA 02155}
\begin{abstract}
Non-equilibrium transport of particles embedded in a liquid crystal
host can, by cooling through a phase transition, be exploited to create
a remarkable variety of structures including shells, foams, and gels.
Due to the complexity of the multicomponent system and protocol-dependent
experimental results, the physical mechanisms behind structure selection
remain only partially understood. Here we formulate a new model coupling
LC physics to a Fokker-Planck equation as is commonly used in studies
of transport. The resulting model allows us to draw an analogy between
the LC-nanocomposite system and chemotaxis, enriching the space of
possible target structures that could be produced. We study the model
in one dimension both analytically and numerically to identify different
parameter regimes where soliton-like pulses of particles ``surf''
the phase boundary or where the interface ``sweeps'' particles from one
domain to another. We also consider an extended model that includes
agglomeration of the particles and observe formation of periodic structures
as a prototypical example of hierarchical self assembly. Results are
compared with experimental observations of transport by isolated phase
boundaries. 
\end{abstract}
\maketitle

\section{Introduction}

Nematic liquid crystal-nanoparticle (NLC-NP) composites are an attractive
union of two very different materials: Submicron colloidal particles possess unique physical, chemical and optical properties due to their small size and high surface area to volume ratio. These particles, when dispersed in a host anisotropic NLC solvent, provide a spatially ordered, externally controllable, and optically active medium. Adjusting one or both components provides
a great degree of control over macroscopic properties such as specific
heat \citep{Vollmer_Hinze_Ullrich_Poon_Cates_Schofield_2005}, shear
and storage modulus \citep{kumar_surface_2020,Petrov_Terentjev_2001,Meeker_Poon_Crain_Terentjev_2000,Anderson_Terentjev_2001},
dielectric anisotropy, and elastic constants\citep{qi_miscibility_2009}.

The liquid crystal component can also guide directed assembly of the
colloidal particles\citep{10.1039/c3sm51628h}. When the system is
cooled from a homogeneous initial state through the host's nematic-isotropic
(NI) phase transition, dispersed nano-\citep{natcomm2019} and
even micron-scale\citep{WestDrag} particles are observed to be transported
by the moving NI phase boundary. The particles are typically coated
with specially designed ligands to promote stability in the host phase and
control aggregate formation \citep{Keshavarz_Riahinasab_Hirst_Stokes_2019,qi_miscibility_2009,tuningnothirst,10.1039/c2sm07091j}.
By adjusting experimental conditions, a number of morphologies can
be produced with structure on length scales that are orders of magnitude
larger than the constituent particles: solid hollow capsules which can be
extracted from the host\citep{hirst2015,hirsttuning,tuningnothirst,natcomm2019},
hierarchical structures formed by aggregation of these hollow capsules in liquid crystal
\citep{devika2022}, as well as networks and solid foams\citep{diestra-cruz_hierarchical_2015,natcomm2019,Dynamicarrest,cleaver_poon_2004}.
Some structures can be formed reversibly and are stabilized by the
host solvent\citep{reversiblechemarxiv,Lesiak, Rodarte_2013}, whereas structures can also be stabilized by taking advantage of ligand-ligand interactions \citep {Keshavarz_Riahinasab_Hirst_Stokes_2019}.

Exploiting the wide variety of morphologies and material properties
possible with these materials may lead to applications in optics \citep{10.1364/ome.5.001469},
device fabrication \citep{10.1889/1.2210802}, medicine \citep{10.1016/j.nantod.2020.101019},
cosmetics \citep{10.4103/0019-5154.96186} and food science \citep{10.1080/10408391003785417}.
Nonetheless, an impediment to adopting and optimizing NLC-NP composites
for particular applications is the lack of a fundamental understanding
of what guides structure selection.

In this paper, we construct a phenomenological model of particle transport
by a moving phase boundary and use our model to provide a unified
understanding of phenomena observed in the formation of NLC-NP composites. We compare the model with experimental high speed video imaging of quantum dot transport near moving phase boundaries.
A number of authors have previously drawn from the theory of dynamic
critical phenomena \citep{hohenberg_halperin_1977} to model transport
processes in LCs\citep{Soule_Lavigne_Reven_Rey_2012,Matsuyama_2008,10.1103/physreve.90.020501,segura-frontiers,reversiblechemarxiv}.
In the critical phenomena framework, the density of particles and
liquid crystal order parameter are described as continuous fields
that evolve in time according to a coupled system of partial differential
equations derived from a free energy.

Here we take a different approach, describing the phase transition
by an Allen-Cahn equation and the particle transport through a Fokker-Planck
equation. We also consider extensions to this model incorporating
irreversible agglomeration of the particles. The resulting theories
are closely related to the Keller-Segel model\citep{Keller_Segel_1970},
providing an analogy between the formation of NLC-NP composites and
chemotaxis. We study numerical solutions of the theory that predict
periodic assembly of particles as observed experimentally \citep{WestDrag,Itahara}.

The paper is organized as follows: In Section \ref{sec:Background},
we review prior experimental results that our model must predict; we then present
our model in Section \ref{sec:Model} and contrast it with prior approaches.
The new formulation provides an analogy between nanoparticle transport
and chemotaxis, which we discuss in Section \ref{sec:Chemotaxis}.
We present analytical and numerical results in Section \ref{sec:Results},
finding two regimes for particle transport including a ``surfing''
scenario where soliton-like perturbations are advected by the interface without disturbing the nanoparticle distribution
and a ``sweeping'' regime where the interface partially or completely clears nanoparticles from the growing
domain. These results are compared to experimental video imaging data.

We also demonstrate, in Section \ref{sec:Solidification},
that if solidification is added into our model hierarchical structure
formation naturally emerges. We draw conclusions in Section \ref{sec:Conclusions}. 

\section{Background}

\label{sec:Background}

A number of experiments have characterized transport of particles
during liquid crystal phase transitions\citep{natcomm2019,Lesiak,reversiblechemarxiv,Dynamicarrest}.
Micron-sized particles show transport of particles influenced by the
nematic director with attraction to and trapping in isotropic domains
\citep{10.1039/c3sm51628h}, and that these effects could be strongly
affected by the particles' anchoring condition \citep{Skarabot_Lokar_Musevic_2013}.
Such particles can be transported by spatially varying order parameters,
either induced by temperature gradients\citep{Kolacz_Konya_Selinger_Wei_2020}
or by dopants\citep{Samitsu_Takanishi_Yamamoto_2010}. With a specific
ligand choice and particle size, logistic recovery of particles back
into nematic domains following initial expulsion has also been observed\citep{reversiblechemarxiv}.

Theoretical work on particle transport in liquid crystals has shown that particles
can travel in either direction with respect to an order parameter
or temperature gradient depending on a competition between thermophoretic
and elastic forces\citep{Kolacz_Konya_Selinger_Wei_2020}. Near an
interface, the force on a particle is $\propto dr$
with $d$ the distance to the interface and $r$ the particle size
and that for a variety of anchoring conditions the free energy is
reduced as the particle approaches the interface\citep{Andrienko_Tasinkevych_Patricio_Teo_da_Gama_2004}.

Structure selection is strongly affected by ligand choice, nanoparticle
concentration, and cooling rate. The ligand affects the particles'
solubility, shifts the $T_{NI}$ transition temperature and determines whether the final structure is stable; surprisingly
the packing density of the particles is not correlated to the shell
radius\citep{tuningnothirst}. Dopants are also known to induce slowing
down of the nematic-isotropic interface propagation speed\citep{Vollmer_Hinze_Poon_Cleaver_Cates_2004}.

Hollow structures were observed to form at all cooling rates and spherical shells were the only
aggregates that form at very high cooling rates \citep{natcomm2019}.
Different ligands lead to shells of different radius and thickness
\citep{tuningnothirst}. Slower cooling and increased nanoparticle
concentration leads to larger shells\citep{natcomm2019}. Periodic
aggregation has been observed by a number of authors\citep{reversiblechemarxiv,WestDrag,Itahara_Tamura_Kubota_Uto_2015}
where the spacing and aggregate size are inversely related to the
cooling rate. Approximately periodic aggregation was observed in \citep{hirsttuning}
with an ODA ligand.

Prior modelling of spherical shell formation includes simple scaling arguments, balancing the
stress on the shell due to the phase boundary with that due to the
nanoparticles. This predicts that the thickness of the shell is inversely
proportional to its radius\citep{tuningnothirst}. Ref. \citep{Atzin_Guzman_Gutierrez_Hirst_Ghosh_2018}
models shell formation using Monte Carlo minimization of an effective
free-energy functional including terms that account for nanoparticle-nanoparticle
volume exclusion, the isotropic-nematic phase transition energy and
nematic elasticity respectively. They find the shell thickness is
set by a competition between the first two terms, with elasticity
controlling the activation barrier for the transition. After formation,
a continuum model with flexible boundaries predicts that the shells
can be deformed by the local nematic and self-organize into chain. This was confirmed experimentally in Ref. \citep{PhysRevE.97.032701}.

Network and foam-like structures have also been created. Networks occur
at slower cooling rates, while foams emerge as cooling rate is increased
in \citep{cleaver_poon_2004}. Other authors find foam structures
at intermediate cooling rates \citep{natcomm2019}. Foam cell size
is inversely related to initial particle concentration \citep{Anderson_Terentjev_Meeker_Crain_Poon_2001}.
At higher cooling rates, the networks can transition to a fern-like structure
with smaller empty cells, higher density of cells, and thinner branches
\citep{Dynamicarrest}. Mesogenic ligands intended to promote solubility
result in a diffuse network with higher nanoparticle concentration
at vertices of the network\citep{hirsttuning}.

Some progress has been made on modelling higher order structures.
Ref. \citep{natcomm2019} developed a Lebwohl-Lasher lattice model
of the liquid crystal coupled to a Cahn-Hilliard model representing
the nanoparticle phase separation, and observed segregation of particles
into domain-like structures with the characteristic size of the domains
varying inversely with the cooling rate. Following the critical phenomena
approach of Hohenberg and Halperin\citep{hohenberg_halperin_1977},
authors in ref. \citep{segura-frontiers} formulate an effective free
energy for the particle density and order parameter and solve the
resulting dynamical equations numerically in two dimensions using
finite differences. We shall discuss this approach in more detail
in \ref{sec:modelc}. Finally, formulating a model of spinodal decomposition
in mixtures of a liquid crystal and colloidal particles, ref, \citep{Matsuyama_2008}
finds many morphologies including fibrous, cell-like or bicontinuous
networks.

\section{Model\label{sec:Model}}

Our model describes the configuration of the system by two spatially
varying and time dependent fields, the density of particles $\rho(\mathbf{x},t)$
and the scalar order parameter $S(\mathbf{x},t)$ of the host liquid
crystal. In places where the system is isotropic, $S=0$. We will
assume that the NLC-NP composite is cooled uniformly with an instantaneous
temperature $T(t)$.

Evolution of the liquid crystal order parameter is given by the equation,
\begin{equation}
\beta\frac{\partial S}{\partial t}=\nabla\cdot\frac{\partial F}{\partial\nabla S}-\frac{\partial F}{\partial S},\label{eq:DynamicalEquationS}
\end{equation}
where $F$ is the free energy of the system and $\beta$ is a transport
coefficient. We use the free energy, 
\begin{equation}
F=\frac{1}{2}\alpha(\nabla S)^{2}+\frac{3}{4}a_{o}\Delta TS^{2}-\frac{1}{4}bS^{3}+\frac{9}{16}cS^{4},\label{eq:FreeEnergy}
\end{equation}
where the first term is an elastic term with parameter $\alpha$ that
penalizes variation in $S$ and the last three terms are the Landau
expansion with associated coefficients $a_{0}$, $b$ and $c$ that
select a preferred value of $S$. The preferred bulk value $S_{0}$
determined by minimizing the Landau terms is a function of $\Delta T=T-T_{0}$
where $T_{0}$ is the temperature below which the isotropic phase
is no longer stable.
Between $0<\Delta T<b^{2}/(27a_{0}c)$ both phases are stable but
the nematic phase has lower energy; between $b^{2}/(27a_{0}c)<\Delta T<b^{2}/(24a_{0}c)$
the isotropic phase has lower energy and for $\Delta T>\Delta T_{c}=b^{2}/(24a_{0}c)$
only the isotropic phase is stable. Substituting Eq. \eqref{eq:FreeEnergy}
into \eqref{eq:DynamicalEquationS} yields an Allen-Cahn-like equation,
\begin{equation}
\beta\frac{\partial S}{\partial t}=\alpha\nabla^{2}S-\frac{3}{2}a_{0}\Delta TS+\frac{3}{4}bS^{2}-\frac{9}{4}cS^{3}.\label{eq:scalarorderparameterevolution}
\end{equation}

We describe the evolution of the particles using a Fokker-Planck equation,
\begin{equation}
\partial_{t}\rho=\mathbf{\nabla}\cdot\left[D\mathbf{\nabla}\rho-\mathbf{f}(S)\rho\right],\label{eq:particleevolution}
\end{equation}
where $D$ is the diffusion constant. The vector quantity $\mathbf{f}(S)$
represents an external force that acts on the particles which we assume
must depend on the phase field $S$. We may construct an expression
for $\mathbf{f}(S)$ from the order parameter $S$ and its derivatives
$\nabla S$ etc and the simplest choice consistent with experimental
studies of particle transport in LCs\citep{Samitsu_Takanishi_Yamamoto_2010}
is, 
\begin{equation}
\mathbf{f}(S)=-\chi\mathbf{\nabla}S,\label{eq:forceansatz}
\end{equation}
where $\chi$ is a parameter that characterizes the degree of forcing. 

The form of Eq. \eqref{eq:forceansatz} suggests an interpretation
for $S$, in that it acts as an effective potential for the nanoparticles.
Depending on the sign of $\chi$, the particles will either tend to
be transported to the nematic domain ($S>0$) or to the isotropic
domain ($S=0$). Close to a phase boundary, the particles
experience a force and are pushed along. Far from a phase boundary
the distribution of nanoparticles relaxes diffusely.

We now discuss a number of simplifying assumptions that we have made
in formulating the above model. We assume that the particles are non-interacting,
except for the diffusive behavior explicitly modeled. We consider
only isotropic diffusion of the nanoparticles, because in the scenario considered here the particles largely remain in the
isotropic phase and are strongly driven by advective forces at the
interface. We neglect flow effects by using an equation of the form
\eqref{eq:DynamicalEquationS}.

Another assumption is that the behavior of the LC is unaffected by
the presence of nanoparticles since Eq. \eqref{eq:scalarorderparameterevolution}
does not depend on $\rho$. Such a coupling between the nanoparticle
concentration and the phase behavior could emerge in a number of ways:
The LC transition temperature itself has been theoretically predicted\citep{10.1039/c0sm01398f,10.1140/epje/i2004-10150-9}
and experimentally shown\citep{natcomm2019,reversiblechemarxiv,Lesiak}
to be modified by the presence of nanoparticles depending on their radius. The interface velocity
might be affected by the assembly of nanoparticles, as suggested in
\citep{cleaver_poon_2004} and modeled in \citep{10.1140/epje/i2004-10150-9}.
Further, in the critical phenomena formalism described below, inclusion
of $S$ in the dynamic equation for $\rho$ implies that $\rho$ should
be included in the dynamical equation for $S$. While such couplings
are likely to be important for determining the nucleation rate\citep{Dynamicarrest}, they can be neglected if the nanoparticle concentration is very dilute and in any case do not appear to be essential to predict the assembly phenomena studied here. We similarly
neglect possible temperature dependence of other quantities such as
$D$ and $\chi$, largely because we are interested in behavior close
to the transition temperature. Finally, by considering a scalar order parameter only,
we cannot capture phenomena such as LC defect formation and interactions
with nanoparticles \citep{10.1103/physreve.61.2831,10.1103/physreve.87.032501}.

\subsection{Nondimensionalization\label{subsec:Nondimensionalization}}

We nondimensionalize Eq. \eqref{eq:DynamicalEquationS} by introducing
a timescale $\tau$, a lengthscale $\xi$ and rearranging,
\begin{equation}
\frac{\partial S}{\partial t}=\alpha'\nabla^{2}S-a_{0}'\Delta TS+b'S^{2}-c'S^{3},\label{eq:dynamicalSnondim}
\end{equation}
with new constants, 
\[
\alpha'=\frac{\alpha\tau}{\xi^{2}\beta},\ a_{0}'=\frac{3}{2}\frac{a_{0}\tau}{\beta},\ b'=\frac{3}{4}\frac{b\tau}{\beta},\ c'=\frac{9}{4}\frac{c\tau}{\beta}.
\]
We similarly nondimensionalize the Fokker-Planck equation, 
\begin{equation}
\frac{\partial\rho}{\partial t}=\nabla\cdot\left[D'\nabla\rho+\chi'\left(\nabla S\right)\rho\right],\label{eq:FPnondim}
\end{equation}
where,
\begin{equation}
D'=\frac{\tau D}{\xi^{2}},\ \chi'=\frac{\tau\chi}{\xi^{2}}.
\label{eq:dchinondim}
\end{equation}
We use values for the material 5CB $a_{0}=0.087\times10^{6}N/m^{2}/K$,
$b=2.13\times10^{6}N/m^{2}$, $c=1.73\times10^{6}N/m^{2}$ and $\alpha\sim1\times10^{-11}N$\citep{stark_2001}.
The transport coefficient $\beta=\gamma_{1}/(9S_{b}^{2})$ is related
to the rotational viscosity $\gamma_{1}=\alpha_{3}-\alpha_{2}=0.0777\ \text{Pa}\cdot s$
from Stewart\cite{stewart2019static} and is approximately constant with temperature\cite{cui1999temperature}.

We may conveniently choose a timescale $\tau$ so that $\beta/\tau=10^{6}N/m^{2}$
cancels the magnitude of the Landau coefficients, implying $\tau\sim10ns$.
We also choose $\xi^{2}$ so that $\alpha'=\frac{\alpha\tau}{\xi^{2}\beta}=1$,
leading to $\xi\sim3nm$. Having nondimensionalized the model, we
will drop the primes henceforth and refer only to the dimensionless
parameters. 

\subsection{Comparison with the critical phenomena framework\label{sec:modelc}}

As noted earlier, a number of authors have previously constructed
models for particle transport drawing upon the theory of dynamic critical
phenomena\citep{Soule_Lavigne_Reven_Rey_2012,Matsuyama_2008,10.1103/physreve.90.020501,segura-frontiers,reversiblechemarxiv}.
To construct such a model, one begins with a free energy expressed
as a function of relevant thermodynamic quantities, here $S$ and
$\rho$. The evolution of these fields is then described by dynamical
equations derived from the free energy. The scalar order parameter
evolves as we already described in \eqref{eq:DynamicalEquationS},
while the conserved particle density must evolve as follows, 
\begin{equation}
\frac{\partial\rho}{\partial t}=\nabla^{2}\left[\frac{\partial f}{\partial\rho}-\mathbf{\nabla}\cdot\frac{\partial f}{\partial\mathbf{\nabla}\rho}\right].\label{modelCrho}
\end{equation}
The combined evolution of a conserved quantity, the total number of
particles, and a non-conserved quantity, the LC order, is referred
to as `Model C' in the classification of Hohenberg and Halperin\citep{hohenberg_halperin_1977}.

Our model is closely related to, but distinct from, this framework.
While the equation for the evolution of $S$ is similar, the alternative
formulation for $\rho$ as a Fokker-Planck equation is more general
because non-conservative forces, i.e. those that cannot be derived
from a free energy, can be included. It also facilitates a rich connection
to other fields of physics that describe transport through Fokker-Planck equations
as we shall momentarily show.

\subsection{Analogy with chemotaxis}

\label{sec:Chemotaxis}

In this section we show that the model formulated above can be mapped
onto the Keller-Segel model\citep{Keller_Segel_1970} of \emph{chemotaxis},
a process by which the motion of autonomous agents, which could be
entire organisms or individual cells, move in response to chemical
cues. Their original paper considered the aggregation of slime molds,
incorporating a feedback loop by which amoebae create reactants that
react to produce a byproduct, acrasin, that serves to attract other
amoebae.

Stated in a general form\citep{Arumugam_Tyagi_2021}, the Keller-Segel
model describes the co-evolution of a population of homogeneous agents
with spatial distribution $p$, 
\begin{equation}
\frac{\partial p}{\partial t}=\nabla\cdot\left[\phi(p,q)\nabla p-\psi(p,q)\nabla p\right]+f(p,q),\label{eq:ksu}
\end{equation}
together with the concentration of a chemical cue $q$, 
\begin{equation}
\frac{\partial q}{\partial t}=d\nabla^{2}q+g(p,q)-h(p,q)q.\label{eq:ksv}
\end{equation}
A number of quantities must be specified to complete the model: $\phi(p,q)$
controls the agents' diffusion, $\psi(p,q)$ advection and $f(p,q)$
reproduction while the functions $g(p,q)$ and $h(p,q)$ specify the
dynamics of the cue.

To make the analogy between our model and the Keller-Segel explicit,
we note that our density of nanoparticles $\rho$ maps onto the density
of agents $p$, while the scalar order parameter $S$ maps onto the
concentration of chemical cue $q$. We see that Eq. \eqref{eq:ksu}
parallels the form of \eqref{eq:particleevolution}. Looking at these
equations term by term, we first note that the diffusion function
$\phi(p,q)$ must be set constant to map onto $D$ in \eqref{eq:particleevolution}.
The last term in Eq. \eqref{eq:ksu} is also straightforward: since
the number of particles is conserved, $f(p,q)$ must be set to zero.

The advection term $\psi(p,q)\nabla p$ in Eq. \eqref{eq:ksu} amounts
to a particular choice of force function in Eq. \eqref{eq:particleevolution}.
The force function used in the K-S model is $\propto\nabla q$, motivated
by assumption that the amoebae are \emph{``sensitive to the relative
acrasin gradient''} \citep{Shaffer_1957,Keller_Segel_1970}. Hence
if $\psi(p,q)\propto p$, the advective term is equivalent to our
choice of force function for the particle transport model Eq. \eqref{eq:forceansatz},
similarly proportional to $\nabla S$.

Now comparing the other pair of parallel equations Eq. \eqref{eq:ksv}
and \eqref{eq:scalarorderparameterevolution}, we see that by an appropriate
choice of polynomials $g(p,q)$ and $h(p,q)$ the two equations can
be made equivalent: the Laplacian term recovers curvature driven dynamics
while terms arising from the Landau expansion amount to a particular
choice of dynamics.

A great benefit of this analogy is that we may in future exploit the
wealth of results about the Keller-Segel model, its extensions, and
numerical techniques to solve it. Indeed, since Keller and Segel's
original paper, their model has been adapted and extended to study
many other chemotactic processes, including hydrodynamic effects
through the Keller-Segel-Navier-Stokes model, logistical growth
models for reproduction, and modulations to the diffusion, to name
a few \citep{Strehlthesis,TysonSternLeVeque2000,Arumugam_Tyagi_2021,Khaled-Abad_Salehi_2021}.

Of particular significance to the present work is that hierarchical self assembly structure formation is also observed in chemotactic systems. Endotheliel cells, for example, have been shown to migrate under chemotactic influence to form vascular structures\cite{ambrosi2004cell,tosin2006mechanics} that remarkably resemble network structures observed in the LC-nanoparticle composites. Such a process can generate a wide range of cellular morphologies including clusters and networks\cite{merks2008contact}. Other examples of morphogenesis rely on chemotaxis-mediated pattern formation\cite{ho2019feather}. Despite the very different physics underlying the two classes of system, the new analogy presented here between phase-transition driven assembly and chemotactic assembly could provide a route to creating new biomimetic structures or even scaffolding new biological structures by exploiting results from the LC-nanocomposite system.

\section{Results\label{sec:Results}}

\begin{figure}
\begin{raggedright}
\includegraphics[width=1\columnwidth]{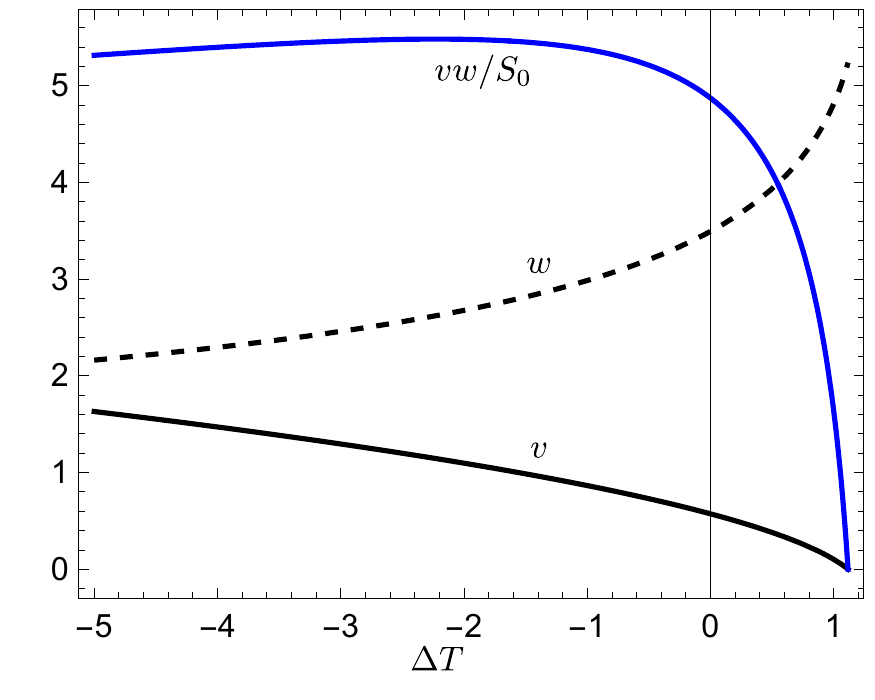}
\par\end{raggedright}
\raggedright{}\caption{\textbf{\label{fig:VWS}} \textbf{Interface velocity $v$ and width
$w$ as a function of temperature.} Also shown is the quantity $vw/S_{0}$,
which determines the critical velocity.}
\end{figure}

\begin{figure}
\begin{raggedright}
\includegraphics[width=1\columnwidth]{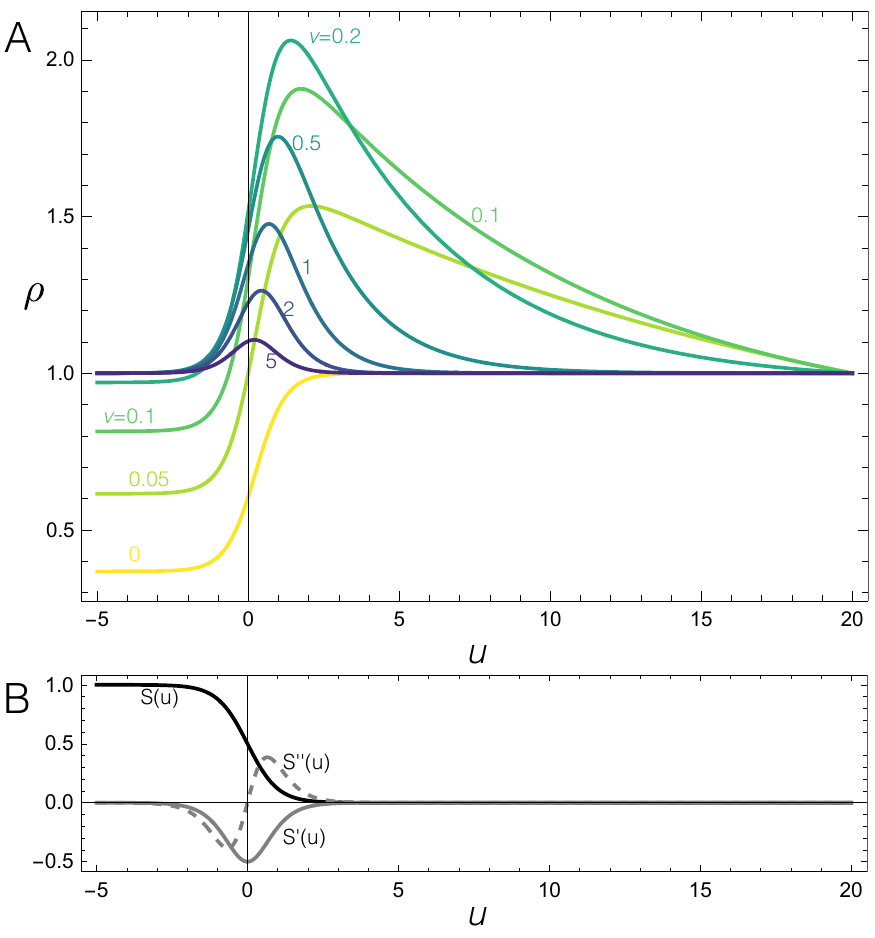}
\par\end{raggedright}
\raggedright{}\caption{\textbf{\label{fig:Analytical}Transition from `surfing' to `sweeping' in nanoparticle distributions guided by a
moving interface.} \textbf{A} Distributions $\rho(u)$ shown as a function of velocity with $w=1$ and parameters chosen so that $v_{c}=1$ and $\frac{\chi S_{0}}{D w^{2}}=1$.  For low interface speeds $v < v_c$, nanoparticles are removed from the nematic domain; for $v > v_c$, a soliton like pulse of nanoparticles `surfs' the phase boundary. \textbf{B} Profile of the corresponding nematic-isotropic interface $S(u)$ and its derivatives. Calculations were performed on a domain $[-L,L]$ with $L=20$ and only the right portion shown.
}
\end{figure}

In this section, we obtain analytical and numerical solutions to our
model Eqs. \eqref{eq:dynamicalSnondim} and \eqref{eq:FPnondim}.
For the purposes of this work, we shall focus on transport and assembly
by a isolated moving interface. Isolated moving interfaces are observed
in the initial stages of structure formation\cite{natcomm2019}, as nematic domains
are nucleated and grow upon cooling. Fluorescence images of the particle
distribution exhibit a band of peak intensity adjacent to the phase
boundary. Such a geometry is one dimensional and will be the subject
of the remainer of this paper. In experimental geometries, domain
growth is typically circular or spherical with a radius $R(r)$ and
hence the nematic-isotropic interface is curved. For simplicity, we
shall neglect curvature which leads to a term $\propto\frac{\alpha}{r}\frac{\partial S}{\partial r}$
in \eqref{eq:dynamicalSnondim} arising from the Laplacian in polar
coordinates; analogous terms arise in \eqref{eq:FPnondim}. Such a
term is negligible for $r\gg1$ and if $R\gg w$ where $w$ is the
width of the interface\cite{bray_1994}.

\subsection{Transport by a moving interface}

To gain initial insight into the character of the solutions of our
model Eqs. \eqref{eq:dynamicalSnondim} and \eqref{eq:FPnondim},
we begin with an idealized scenario: A single interface between semi-infinite
nematic and isotropic domains which moves at a constant velocity $v$,
for which an analytical solution to \eqref{eq:dynamicalSnondim} exists.
In the comoving frame $u=x-vt$, Eq. \eqref{eq:dynamicalSnondim}
becomes,
\begin{equation}
\alpha\frac{d^{2}S}{du^{2}}+v\frac{dS}{du}-a_{0}\Delta TS+bS^{2}-cS^{3}=0\label{eq:dynamicalScomoving}
\end{equation}
with a well-studied solution\cite{Popa-Nita_Sluckin_1996}, 
\begin{equation}
S=S_{0}\left[1-\tanh(u/w)\right].\label{eq:Scomoving}
\end{equation}
which represents an interface of width $w$ about $u=0$ where the
order parameter interpolates from the nematic phase $S=S_{0}$ for
$u\ll0$ to the isotropic phase $S=0$ for $u\gg0$.

Inserting Eq. \eqref{eq:Scomoving} into Eq. \eqref{eq:dynamicalScomoving}
and solving for $v$ and $w$ gives, 
\begin{equation}
v=\frac{bS_{0}-3a_{0}\Delta T}{\sqrt{2}\sqrt{bS_{0}-a_{0}\Delta T}},w=\frac{2\sqrt{2 \alpha}}{\sqrt{bS_{0}-a_{0}\Delta T}},\label{eq:vw}
\end{equation}
At $\Delta T_{eq}$, the point at which the isotropic and nematic
phase have equal energy, the interface velocity is zero. If the temperature
is reduced, the interface width decreases, while the velocity of the
interface increases as shown in Fig. \ref{fig:VWS}. 

Transforming the Fokker-Planck equation to the comoving frame yields,
\begin{equation}
\frac{\partial\rho}{\partial t}=D\frac{\partial^{2}\rho}{\partial u^{2}}+\left(v+\chi\frac{\partial S}{\partial u}\right)\frac{\partial\rho}{\partial u}+\chi\frac{\partial^{2}S}{\partial u^{2}}\rho.\label{eq:FPcomoving}
\end{equation}
We seek equilibrium solutions $\frac{\partial\rho}{\partial t}=0$,
and hence \eqref{eq:FPcomoving} becomes a second order homogenous
ODE with spatially varying coefficients,
\begin{equation}
D\frac{d^{2}\rho}{du^{2}}+A(u)\frac{d\rho}{du}+B(u)\frac{d^{2}S}{du^{2}}\rho=0,\label{eq:FPcomovingtimeindep}
\end{equation}
where the coefficients are found by inserting the solution \eqref{eq:Scomoving}
into \eqref{eq:FPcomovingtimeindep}, 
\begin{align}
A(u) & =\left(v-\frac{S_{0}\chi}{2w}\text{sech}^{2}(u/w)\right),\label{eq:Spatiallyvaryingcoefficients}\\
B(u) & =\frac{S_{0}\chi}{w^{2}}\text{sech}^{2}(u/w)\tanh(u/w).
\end{align}
Since all parameters are positive, $B(u)$ has a zero at $u=0$, while
$A(u)$ has no zeros if $v>\frac{S_{0}\chi}{2w}$ and two placed at
$u=\pm\text{arccosh}\left(\sqrt{S_{0}\chi/2vw}\right)$ if $0<v<\frac{S_{0}\chi}{2w}$.
We therefore identify a critical velocity,
\begin{equation}
v_{c}=\frac{S_{0}\chi}{2w},\label{eq:vcritical}
\end{equation}
and identify two important numbers,
\begin{equation}
\zeta=\left(v-v_{c}\right)/D,\ \eta=\frac{\chi S_{0}}{Dw^{2}},\label{eq:numbers}
\end{equation}
that characterize the solution. The quantity $vw/S_{0}$ is shown
in Fig. \ref{fig:VWS}; if $vw/S_{0}>\chi$ then $\zeta$ is positive. 

We now seek solutions consistent with the boundary conditions $\rho(+L)=\rho_{0}$
and $\rho'(-L)=0$ for some $L\gg w$. Such a solution for $v=0$
is,
\begin{equation}
\rho=\rho_{0}\exp\left(\frac{S_{0}\chi}{2D}\left[\tanh(u/w)-1\right]\right),\label{eq:rhozero}
\end{equation}
which interpolates between a lower concentration in the $u<0$ nematic
region and a higher concentration $u>0$ in the isotropic region.
For finite $v$, we solve \eqref{eq:FPcomovingtimeindep} numerically
on a finite domain $[-L,L]$ for various values of $v$ and display
the results in Fig. \ref{fig:Analytical} together with the solution
\eqref{eq:Scomoving} and its derivatives. For larger values of $v\gtrsim v_{c}$,
solutions resemble a soliton-like pulse around $u=0$ and the solution
approaches $\rho_{0}$ on the left hand boundary. As $v$ is reduced
the pulse becomes larger and increasingly asymmetric. Decreasing $v$
further, around $v\lesssim\frac{S_{0}\chi}{2L}$ the solution becomes
affected by the right hand boundary condition and gradually deforms
to agree with the $v=0$ solution. We therefore identify two regimes:
a ``surfing'' regime where a pulse of particles is advected by the
interface with the background undisturbed at $v\gtrsim v_{c}$ and
a ``sweeping'' regime $0\leq v\ll v_{c}$ whereby the interface
partially or completely clears the nematic domain. 

We also observe that the width of the particle distribution can be
much greater than the width of the phase boundary as is consistent
with experiments. For $u\gg0$, the coefficients in Eq. \eqref{eq:FPcomovingtimeindep}
approach constant values $A(u)\to v/D$ and $B(u)\to0$; hence the
leading edge of the solution is of the form,
\begin{equation}
\rho\sim\exp(-uv/D),\label{eq:rholeadingedge}
\end{equation}
and we identify a characteristic decay length $\lambda=D/v$. 

The semi-infinite geometry considered in this section is sufficiently
tractable analytically to enable us to identify qualitative features
of particle distributions driven by a moving interface. However, in
some of the experiments described previously in Section \ref{sec:Background}
the temperature is changing with time. Hence, because the interface
velocity $v$, the domain width $w$, and the order parameter $S_{0}$
are all functions of temperature as discussed above in Section \ref{sec:Model},
these quantities are all potentially functions of time as well as
the system is cooled.

\subsection{Transport by a growing domain}

\begin{figure*}
\begin{centering}
\includegraphics[width=1\textwidth]{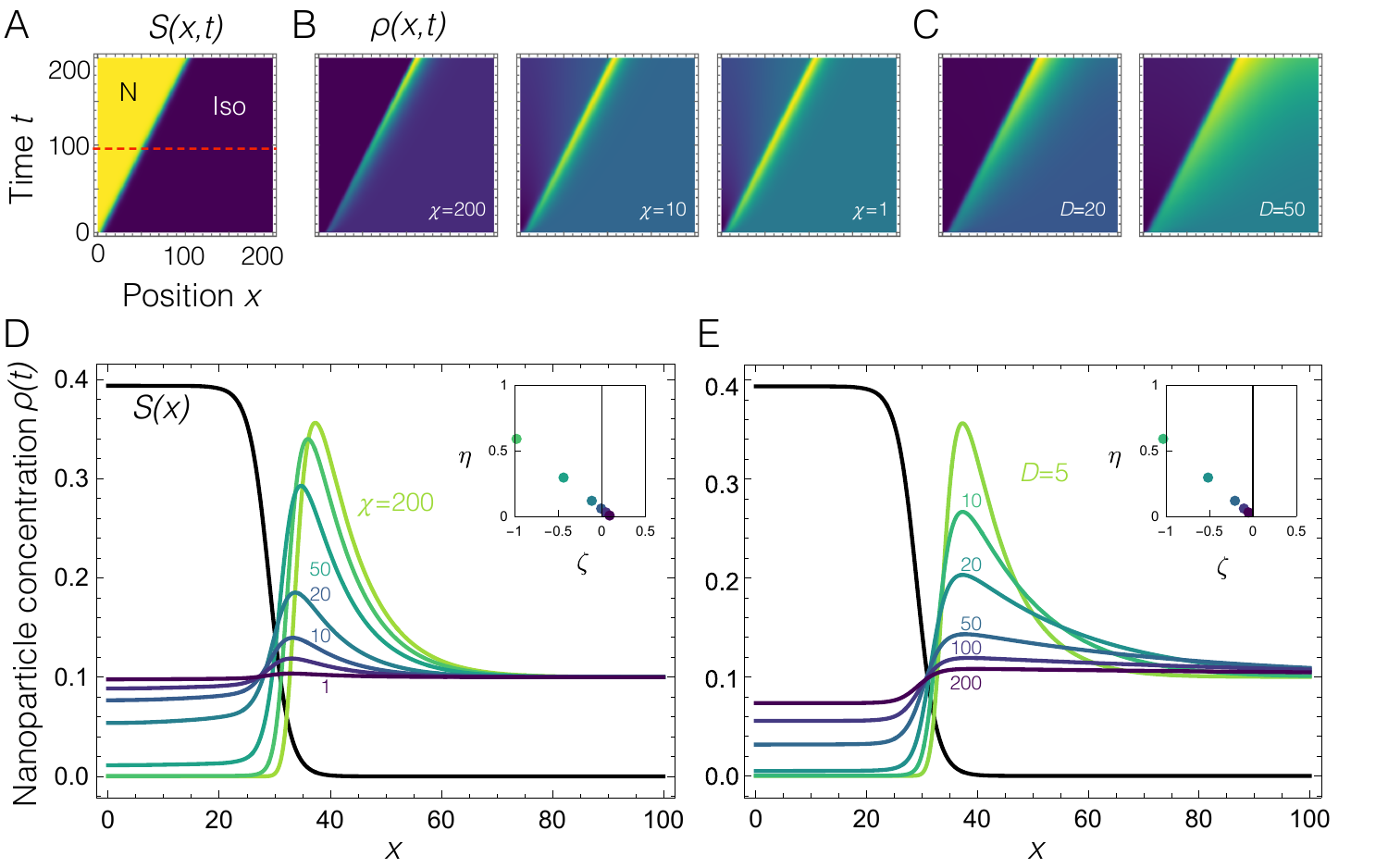}
\par\end{centering}
\raggedright{}\caption{\textbf{\label{fig:Domain}Effect of varying forcing and diffusion constants. A} Kymograph of growing nematic domain with order
parameter $S(x,t)$ at fixed temperature $\Delta T=0.2$. Red dashed
line indicates the cross section visualized in D, E. \textbf{B} Corresponding
particle distribution $\rho(x,t)$ kymographs for $D=5$ and $\chi=200,10,1$.
\textbf{C} Particle distributions fixing $\chi=200$ and $D=20,50$.
\textbf{D} Snapshots at $t=100$ of the interface $S(x)$ (black line)
and particle distribution $\rho(x)$ (color lines) for $D=5$ and
varying $\chi$. \textbf{E} Snapshots at $t=100$ for $\chi=200$
with varying $D$ . \emph{Insets:} Corresponding parameters $\zeta$,
$\eta$ for $D$, $\chi$. All solutions were solved on a domain $[-L,L]$
with $L=200$; only the $x>0$ region is plotted. }
\end{figure*}

The geometry considered in the preceeding section, semi-infinite nematic
and isotropic domains, is quite different from the experimental situation
where a spatially uniform particle distribution that is disrupted
by the nucleation of a nematic domain. We therefore turn to a more
realistic scenario by modeling the behavior of $S$ and $\rho$ \emph{in
the rest frame} with an initial nematic domain seeded around $x=0$
at $t=0$,
\begin{equation}
S(x,0)=S_{0}\exp\left[-\left(\frac{x}{w_{0}}\right)^{2}/2\right],
\label{eq:Sinit}
\end{equation}
where we choose values of $S_{0}$ and $w_{0}$ from \eqref{eq:S0temp}
and \eqref{eq:vw} consistent with an initial temperature $\Delta T$
and use material parameters given in subsection \eqref{subsec:Nondimensionalization}.
The initial condition on $\rho$ is a constant $\rho=\rho_{0}$. We
solve Eqs. \eqref{eq:dynamicalSnondim} and \eqref{eq:FPnondim} in
\texttt{Mathematica} using the Method of Lines on a domain $x\in[-L,L]$
and for $t\in[0,T_{0}]$ with periodic boundary conditions. We choose
$L=200$ and $t=200$ corresponding to a physical domain of $\sim1.2\mu m$
and a time of $\sim2\mu s$.

We first study solutions at a fixed temperature and choose $\Delta T=0.2$,
which is in the region where both nematic and isotropic phase are
stable, but with the nematic phase energetically favored. We examined
solutions for other temperatures $\Delta T<\Delta T_{c}$, the limit
of stability of the nematic phase, and find similar results. 

In Fig. \ref{fig:Domain}A, we display the spatial and temporal evolution
of $S$ as a \emph{kymograph} as might be conveniently extracted from
microscopy data; note only one half $x>0$ of the solution is shown.
The nematic domain grows into the surrounding isotropic phase at constant
velocity, as expected at fixed temperature. Corresponding kymographs
for $\rho$ are displayed in Fig. \ref{fig:Domain}B and C as a function
of the physically intuitive experimental parameters $D$ and $\chi$.
Variation of these parameters could be achieved, for example, by adjusting
the particle size to change the diffusion constant or the ligand coating
the nanoparticles to change $\chi$ as was suggested in \citep{natcomm2019}.

Kymographs showing the effect of varying $\chi$ with fixed $D=5$
are shown in \ref{fig:Domain}B, while in Fig. \ref{fig:Domain}C
we fix $\chi=200$ and vary $D$. For large $\chi$, the nanoparticle
distribution is rapidly depleted from the nematic phase, forms a concentrated
peak in the isotropic domain near the phase boundary and moves with
it as the nematic grows into the isotropic. Such assembly and advective
transport by the phase boundary has been observed experimentally\citep{Dynamicarrest}.

Snapshots of particle distributions at a single moment in time, but
with varying $\chi$ and $D$ are shown in Fig. \ref{fig:Domain}E,
F. In Fig. \ref{fig:Domain}D, we vary $\chi$ with $D=5$
while in Fig. \ref{fig:Domain}E $\chi=200$ and $D$ is varied.
In the insets of each figure we show corresponding values of $\zeta=\left(v-v_{c}\right)/D$
and $\eta=\chi S_{0}/(Dw^{2})$. As predicted from the idealized model
in the previous section, we see complete clearing of the nematic domain
for $\zeta\ll0$, the ``sweeping'' regime, which is the case for
large $\chi$ or small $D$. As $\zeta$ approaches $0$, which happens
if $\chi$ is sufficiently small or $D$ sufficiently large, incomplete
clearing of the nematic domain occurs, a result that has also been
observed experimentally\citep{reversiblechemarxiv}. Also in agreement
with our analysis of the single interface model, the width of the
distribution is proportional to $D$ and not affected by $\chi$. Both parameters affect the overall shape, especially the skewness, of the distribution. 

\begin{figure}
\begin{centering}
\includegraphics[width=1\columnwidth]{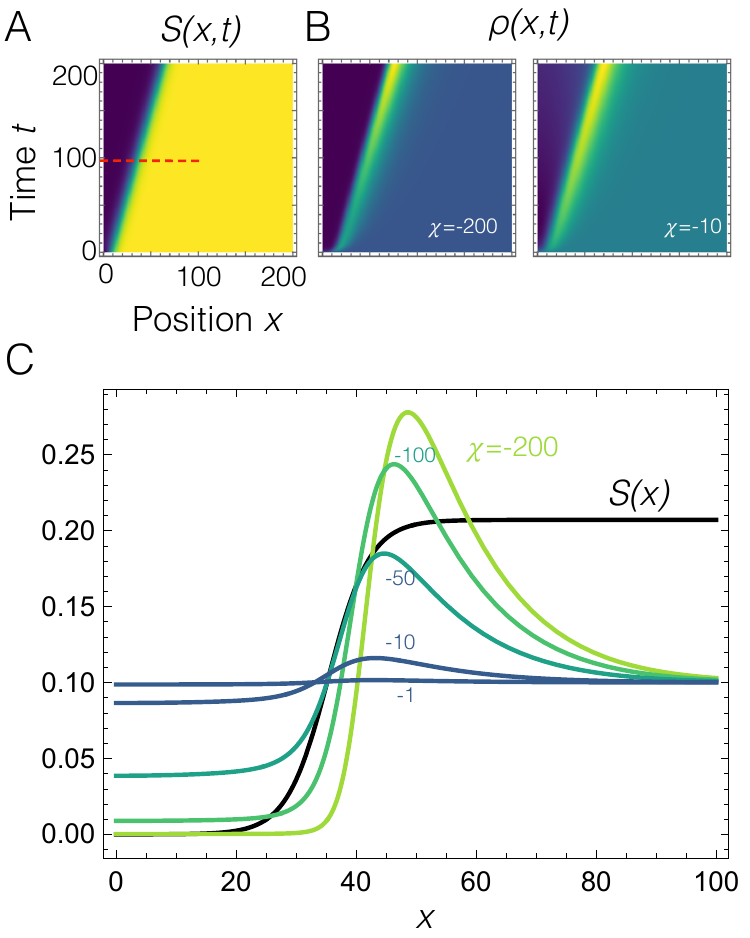}
\par\end{centering}
\raggedright{}\caption{\textbf{\label{fig:Reciprocal}Reciprocal scenario where nanoparticles are guided by a
growing \emph{isotropic} domain. A }Kymograph with
order parameter $S(x,t)$ at fixed temperature $\Delta T=\Delta T_{c}-10^{-4}$.
Red dashed line indicates the cross section visualized in C. \textbf{B}
Corresponding particle distribution $\rho(x,t)$ kymographs for $D=5$
and $\chi=-200,-10$. \textbf{C} Snapshots at $t=100$ of the interface
$S(x)$ (black line) and particle distribution $\rho(x)$ (color lines)
for $D=5$ and varying $\chi$. All solutions were solved on a domain
$[-L,L]$ with $L=200$; only a subinterval of the domain is shown
for clarity.}
\end{figure}

Finally, we consider a reciprocal scenario where an isotropic domain
grows into a nematic domain on heating. Such a situation obviously
occurs only if the temperature is in the coexistence region. As an
illustration, we perform simulations starting from an initial nucleated
isotropic configuration $S=S_{0}\left[1-\exp(-x^{2}/2w_{0}^{2})\right]$
at $\delta T=10^{-4}$ just below $\Delta T_{c}=b^{2}/24a_{0}c$ ($\approx1.25598$
with the Landau coefficients used) to maximize the interface velocity.
The corresponding order parameter distribution is shown in Fig. \ref{fig:Reciprocal}A.
In this situation, if the sign of $\chi$ is also reversed, particles
are driven into the nematic domain from the isotropic as shown in
Fig. \ref{fig:Reciprocal}B and experimentally reported in \citep{Lesiak}.
A similar surfing/sweeping transition occurs as a function of $\chi$
as shown through the snapshots of $\rho$ at a single timepoint in
Fig. \ref{fig:Reciprocal}C.

Hence, by choosing suitable values of the parameters, our model can
account for and unites multiple observed transport scenarios by moving
nematic-isotropic interfaces.



\subsection{Solidification\label{sec:Solidification}}

\begin{figure}
\begin{centering}
\includegraphics[width=1\columnwidth]{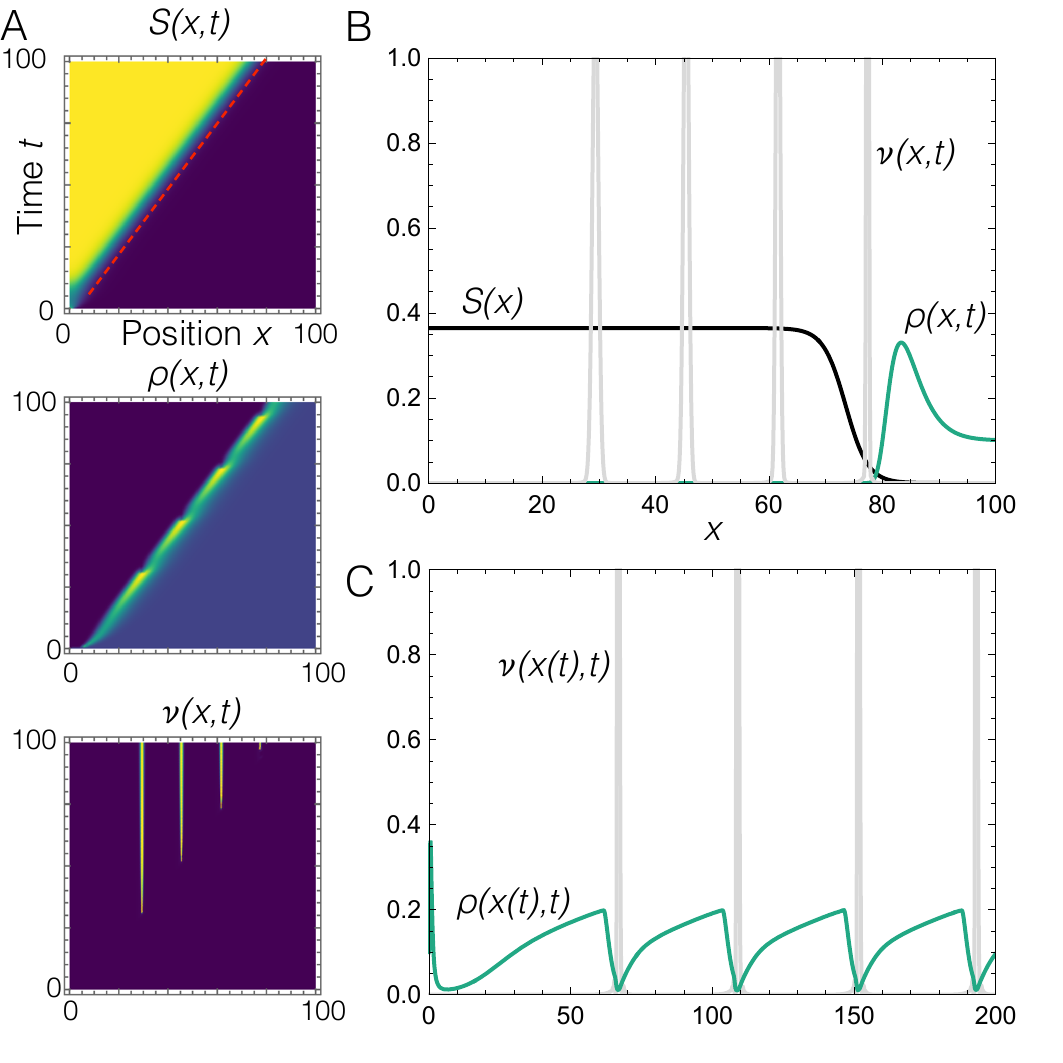}
\par\end{centering}
\raggedright{}\caption{\textbf{\label{fig:Solidification}Hierarchical assembly due to solidification.
A }Kymographs of assemby process showing growing nematic domain with
order parameter $S(x,t)$, mobile particle distribution $\rho(x,t)$
and solid particle distribution $\nu(x,t$). Red dashed line indicates
the cross section visualized in C. \textbf{B} Snapshot at $t=100$
of the system showing periodically assembled bands.\textbf{ C} Cross
section of $\rho$ and $\nu$ along a contour $x=4+vt$ showing periodic
aggregation and solidification. Solution was evaluated on a domain
$[-L,L]$ with $L=200$; only a subinterval of the domain is shown
for clarity.}
\end{figure}

In this section, we create an illustrative example of hierarchical
assembly by constructing an extension of the model developed in Section
\ref{sec:Model} that includes solidification i.e. the irreversible
assembly of the particles. Experimentally, reversible assembly has
been observed \citep{reversiblechemarxiv,Vollmer_Hinze_Ullrich_Poon_Cates_Schofield_2005};
here we focus on modelling situations where the assembly is irreversible
\citep{hirst2015,natcomm2019}, at least on the timescales of the experiment.
We assume that such assembly occurs on a timescale much shorter than
it takes for particles to move a diffusion length.

To do so we introduce a new field $\nu(x,t)$ that describes the distribution
of particles that are no longer mobile. Particles are added to the
assembly if the local particle density $\rho+\nu$ exceeds a critical
density $\rho_{*}$, which is achieved by a rate function,

\begin{equation}
R(\rho,\nu)=\rho\Gamma H(\rho+\nu-\rho_{*})
\end{equation}
where $\Gamma$ is the overall rate of the reaction and $H(z)$ is
the Heaviside step function. For numerical conditioning, we use a
continuous approximation of $H(z)$,
\begin{equation}
H(z)\sim\left[1-\exp(-kz)\right]^{-1},\label{eq:Heavisideapprox}
\end{equation}
with a finite value $k\gg1$.

The Fokker-Planck equation must be modified to include loss of particles
to the solid,

\begin{equation}
\frac{\partial\rho}{\partial t}=\nabla\cdot\left[D\nabla\rho+\chi\left(\nabla S\right)\rho\right]-R(\rho,\nu),\label{eq:nondimrhowitha}
\end{equation}
and a new equation for $\nu$, the immobile particle density, introduced,

\begin{equation}
\frac{\partial\nu}{\partial t}=\epsilon\nabla^{2}\nu+R(\rho,\nu),\label{eq:aevolve}
\end{equation}
The first term in \eqref{eq:aevolve} promotes relaxation of the solid
and is necessary for numerical stability of the overall system \eqref{eq:dynamicalSnondim},
\eqref{eq:FPnondim} and \eqref{eq:aevolve}. Here, we use a small
value for $\epsilon$ to imply very slow relaxation.

In Fig. \ref{fig:Solidification}A we display kymographs showing the
results of a typical simulation. We choose $\chi$ and $D$ so that the simulation is deep in the ``sweeping'' regime whereby particles are completely cleared from the nematic domain by the interface and fix temperature so that the interface
width and velocity are constant. As the nematic domain grows, particles
slowly accumulate ahead of it, and eventually exceed the critical
density. At this point, the excess particles solidify in a narrow
region, returning $\rho$ to lower than its critical value; the process
is therefore reset and repeats as the interface continues to proceed.
On the kymograph for $\nu$, as well as a snapshot of the solution
at $t=T_{0}$, Fig. \ref{fig:Solidification}B, we see the narrow
regularly spaced bands of solid. The repeated process of accumulation
of $\rho$ and solidification is more clearly visible on a plot, Fig.
\ref{fig:Solidification}C, of $\rho$ and $\nu$ at a single co-moving
point $x=x_{0}+vt$ as a function of time. 

\subsection{Comparison with experiment}

\begin{figure*}
\begin{centering}
\includegraphics{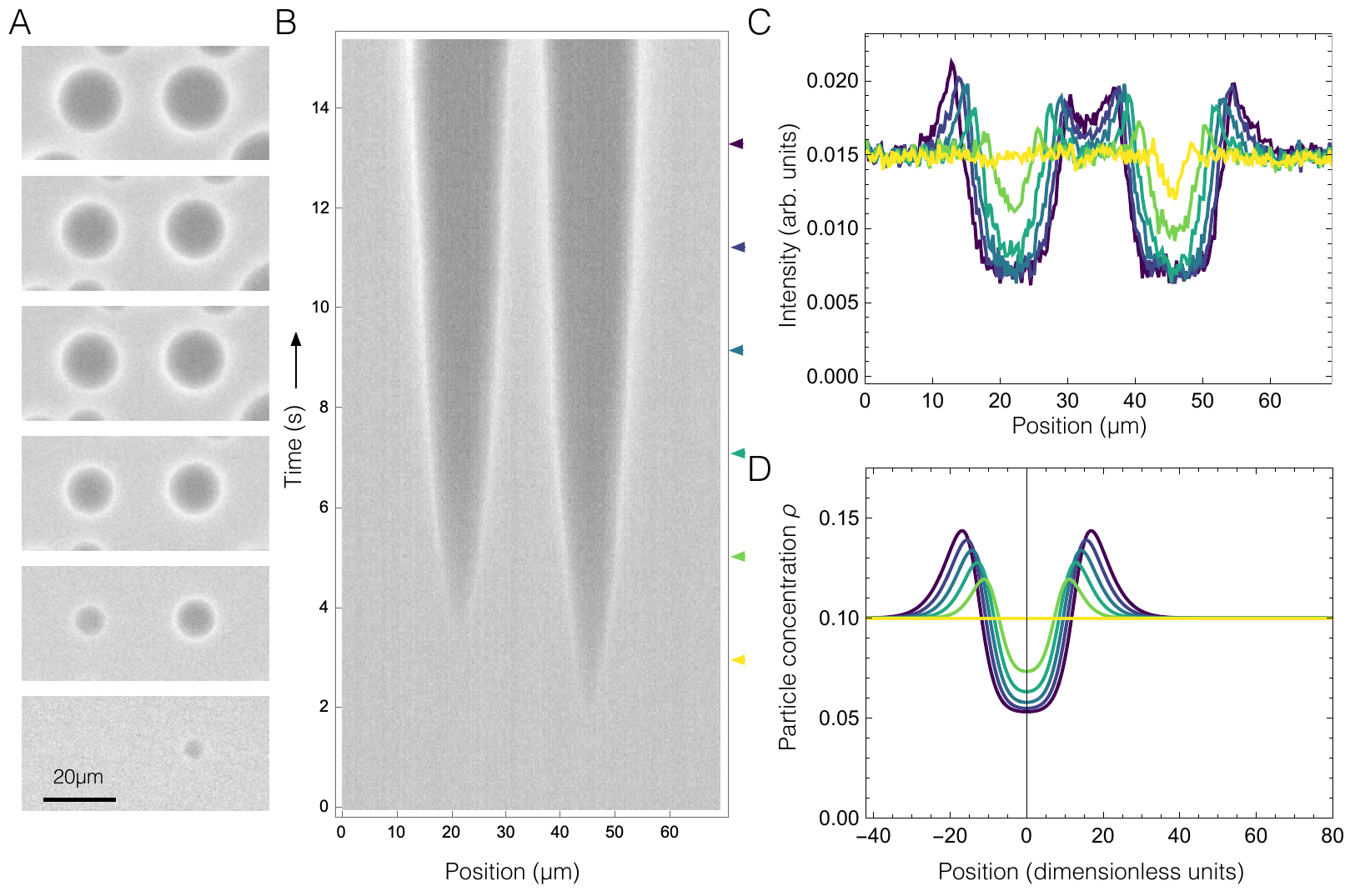}
\par\end{centering}
\raggedright{}\caption{\textbf{\label{fig:Expt}Fluorescence microscope imaging of nanoparticle concentration during nematic domain nucleation and growth events.} \textbf{A} Snapshots of two relatively isolated domains at different timepoints. \textbf{B} Kymograph produced by slicing along the center of the domains. Arrows indicate time points shown in panels A and C. \textbf{C} Selected intensity traces at depicted time points. \textbf{D} Modelled evolution of the nanoparticle distribution.}
\end{figure*}

To test the predictions of the model, we prepared and imaged an experimental
realization of nanoparticle transport guided by a nematic-isotropic
phase transition as follows: We prepared a nanocomposite mixture following our reported protocol \citep{natcomm2019} where a 7.6 mg/mL toluene solution of CdSe/ZnS quantum dots coated with L1 ligand\citep{Keshavarz_Riahinasab_Hirst_Stokes_2019, Riahinasab_2017} is dispersed in 5CB to yield a 0.75\% (w/w) quantum dot solution in 5CB after toluene removal. This mixture was kept in an incubator at 50 $^{\circ}$C, ensuring
that the liquid crystal host remained in the isotropic phase prior
to the controlled phase transition. Inside the incubator, 2.0 $\mu$L of the nanocomposite mixture was pipetted onto a clean glass microscope
slide and covered with a cover glass; A Kapton film spacer was used
to ensure a liquid crystal film thickness of ~25 $\mu$m. The glass
slide was sealed with UV glue and then removed from the incubator
and quickly transferred to a temperature control stage (Linkam TMS94
\& LTS 350, United Kingdom) held at 45 $^{\circ}$C. To record videos
of quantum dot distribution through the phase transition, the stage
temperature was then decreased at a rate of 1$^{\circ}$C per minute
with aid of an in-house designed liquid nitrogen cooled air system.
The phase transition was recorded using a Leica DM2500 LED fluorescence
microscope, I3 fluorescence filter cube, and Phantom camera VEO 410L.
Videos were recorded at various frame rates (24-200 frames per second)
using either a 20X or 40X magnification objective lens.

A sequence of snapshots at different time points for a cropped region
of interest from a representative video are displayed in Fig. \ref{fig:Expt}A; the corresponding movie is provided as Supplementary Information.
In this example, two relatively isolated domains are observed to nucleate
and grow over a timescale of several seconds. A corona-like intensity
maximum is observed around the growing domains, as was seen in \citep{natcomm2019}.
A slice through the image stack going through the horizontal centerline
of the domains is displayed as an $(x,t)$ kymograph in Fig. \ref{fig:Expt}B, showing the growth and movement of the particle corona as the domains
grow. While the domain on the right nucleates slightly earlier, the
growth and evolution of these two domains (and many others from other
regions of interest) are very similar. Selected intensity profiles
at highlighted time points are then shown in Fig. \ref{fig:Expt}
C; these have been averaged over 5 adjacent frames to improve the
signal to noise ratio. 

The time resolved intensity profiles show partial sweeping of nanoparticle
concentration from the growing nematic domain, with a characteristic
peak that grows and is advected along with the domain boundary. The
shape of the peak strongly resembles that observed in our model, and
hence we display snapshots from a simulation in Fig. \ref{fig:Expt}
D. To create these, we used Landau parameters for 5CB as described
in subsection \ref{subsec:Nondimensionalization} and used $\Delta T=0.1$ estimated from the observed timescale
of the growth together with the known cooling rate. We adjusted $\chi=4$
and $D=0.6$ to approximately match the shape of the traces. Other
choices of parameters are possible to reproduce the data, and careful
independent measurements of the viscosity and/or diffusion constant
would be needed to find a unique fit. Nonetheless, the shape, behavior
and growth of the nanoparticle concentration predicted by the model
is certainly observed in this data. Surprisingly, though, the interface
velocity does not increase with time, as would be expected from the
Landau model, but rather decreases. This could be due to intradomain
interactions, or because the nanoparticle concentration strongly affects
the transition temperature of the composite in the vicinity of the phase boundary
and hence slows down the interface. 
\section{Conclusions\label{sec:Conclusions}}

In this work, we have developed a model of particle transport by a
moving phase boundary that couples an Allen-Cahn like equation arising
from the dynamics of the nematic order parameter to a Fokker-Planck
equation describing particle transport. Our model is related to, but
departs from, the `Model C' approach previously used to describe from
the framework of critical phenomena. Formulating particle transport
using the Fokker-Planck approach allows for forces that cannot be
expressed as derivatives of a free energy, and also enables us to
connect with the vast literature on stochastic transport that leverages
this framework. As an illustration, we draw a new correspondence between
our model and the Keller-Segel model of chemotaxis for living systems.
This analogy suggests a number of new possible structures that might
be created by the LC system, and offers the possibility of using advanced
numerical techniques developed to solve such equations. 

As an illustration of our model, we consider particle transport by
an isolated phase boundary as is visible in the early stages of structure
formation. By performing a careful analysis of this restricted scenario
here, we shed new light on possible mechanisms guiding assembly of
more complex structures, such as the gels, shells and foams experimentally
observed. With one model, we reproduce numerous disparate experimental
results including a transition between `sweeping' particles from a
growing nematic domain and soliton-like groups of particles `surfing'
the interface. The interface velocity, width and order parameter,
which are all temperature dependent quantities, play a key role in
determining which regime the system is in. 

By incorporating the possibility of solidification once a critical
density is reached, we are able to produce regularly-spaced solid
deposits, a one dimensional version of the hierarchical structures
observed and with features on lengthscales far greater than the nanoparticle
size or the interface width. In future work, we will apply this model
to higher dimensional geometries to resolve the many outstanding questions
about structure selection. 
\begin{acknowledgments}
\emph{TS and TJA designed the theoretical model and performed numerical simulations with input from CJ. JO, AW and LSH designed the experimental realization. ICR and BJS provided ligand L1. JO and TJA analyzed the data. All authors contributed to prepare the manuscript.
This material is based upon work supported by the National Science Foundation under Grant No. DMR-2104575. }
\end{acknowledgments}

\bibliographystyle{apsrev4-2}
\bibliography{bibliography}

\end{document}